# ADAPTATIVE MESH REFINEMENT IN NUMERICAL RELATIVITY[*]

*JOAN MASSÓ*[1,2], *EDWARD SEIDEL*[1], *PAUL WALKER*[1]

[1] *National Center for Supercomputer Applications, 605 East Springfield Avenue, Champaign, IL 61280, USA*

[2] *Departament de Física, Universitat de les Illes Balears, E-07071 Palma de Mallorca, Spain*

ABSTRACT

We discuss the use of Adaptative Mesh Refinement (AMR) techniques in dynamical black hole spacetimes. We compare results between traditional fixed grid methods and a new AMR application for the 1-D Schwarzschild case.

## 1. Introduction

AMR was introduced in the last decade[1] as a very promising technique for numerical treatment of partial differential equations involving strong dynamic ranges. The basis of AMR methods begins by defining a coarse mesh that covers the entire computational domain. Refined grids of higher resolution are added to regions of the domain where additional resolution is required. This process of adding finer and finer meshes continues to some prespecified level of accuracy. In addition, this hierarchical structure is dynamic so that the algorithms are capable of adapting themselves to arbitrary problems by automatically refining and moving meshes to resolve small scale features as they develop and evolve. The result is a tremendous savings of computer memory and a reduction in execution time over large fine grid simulations.

Unfortunately, until recently, AMR has not had major impact in the field of Numerical Relativity. The work by Choptuik[2,3] can be considered pioneering and his recent discovery of critical phenomena in scalar field collapse[3] would not have been possible without the use of AMR techniques.

Black hole numerical spacetimes lead to extreme dynamic ranges in length and time, making it difficult to maintain accuracy and stability for long periods. In Figs. 1a-b we show the results of evolving a dynamically sliced Schwarzschild spacetime with a standard 1D code developed by Bernstein, Hobill and Smarr.[4] The lack of resolution to resolve accurately the growth of the peak in the radial metric function A (see Ref. 4 for details) eventually causes the code to crash. Increasing the resolution of the grid allows longer and more accurate evolutions (note the reduced error of the apparent horizon mass) but does not avoid the final fate, as the growth of the peak is unbounded (but should always be smooth).

In these proceedings, we investigate the use of AMR techniques to obtain more accurate evolutions of black hole spacetimes, as they will provide the necessary resolution in the sharp regions. This does not mean that AMR is the best way to solve this unphysical problem. In this case, AMR could be considered the "Computer Sci-

---

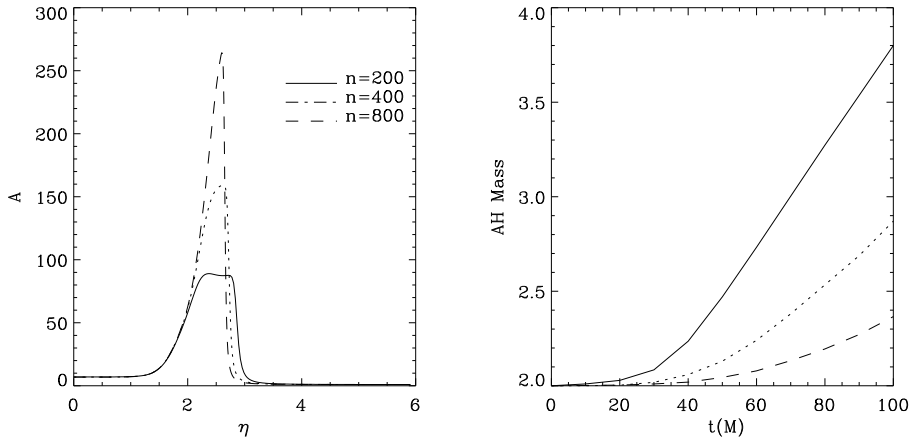

Figure 1: BHS results for different fixed grid sizes. (a) Radial metric component $A$ at $t = 100M$. (b) Time evolution of the apparent horizon mass, which should remain constant $M = 2$.

ence" approach to the problem. The "Relativistic" approach would use the special causal nature of the horizon to excise the region of sharp gradients; this is the idea behind the *Apparent Horizon Boundary Condition* techniques and recent progress is reported in these proceedings.[5] The two approaches are not by any means incompatible and, in any case, there is no penalty for using AMR in all problems as it will be automatically used whenever and wherever is necessary.

Developing an AMR application requires sophisticated data structure management and software engineering. This is probably one of the reasons for the small number of current applications. Despite this, AMR can be developed as a very general "black box" toolkit that can then be applied to any given "solver". Our AMR application has been developed with this approach in mind and can be applied, in principle, to any system of equations in one dimension. For this paper, we converted the BHS code to a subroutine that performs 1 time step on a given grid and it is called by the AMR controller program at each level of the grid hierarchy. Here we will not discuss the details of the AMR toolkit. Instead, we refer the reader to the excellent introduction in Ref. 2 (and references therein).

## 2. Results

In Fig. 2a we show the excellent results obtained with our AMR application for the same case as in Fig. 1a. We use only 3 levels of refinement (and each level refines a factor 4), with a coarse grid (level 0) of only 60 points. The structure of the grid levels is shown in Fig. 1b, where we can see that grid levels are added in the region of the peak growth. This means that AMR allows us a virtual resolution of almost 4000 points. The mass of the apparent horizon is conserved with the equivalent accuracy of this grid. Achieving this results with the BHS code would take around 12 hours of CPU time (on a given workstation) and more than 1 Megabyte of memory. With AMR, it takes only 25 minutes and less than 100K of memory. We are gaining at least a factor 30 in time and 10 in memory. Using more levels of refinement greatly increases this gain.

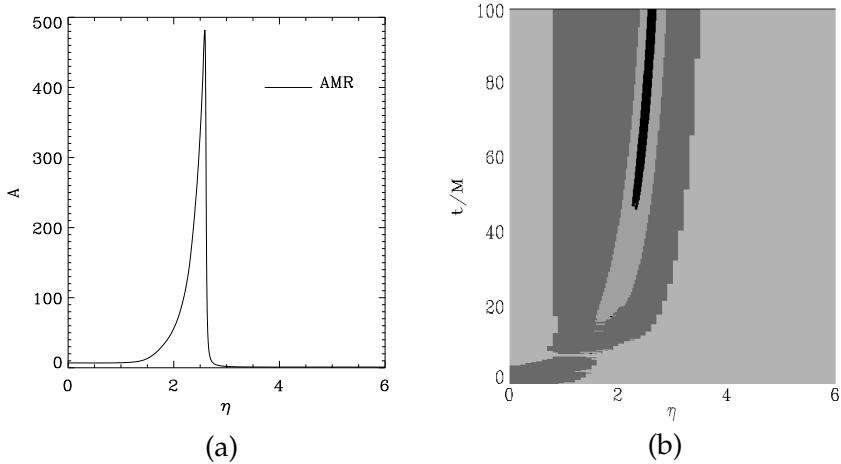

Figure 2: AMR results: (a) Radial metric component $A$ at $t = 100M$. (b) AMR Grid Structure.

These excelent results show the feasibility of using AMR techniques for solving the sharp gradient regions of traditional numerical relativity codes. The extension of their use to 3-D cases will prove challenging.

## Acknowledgements

We thank Pete Anninos and Matt Choptuik for useful discussions. We acknowledge the support of NCSA and NSF grants PHY94-07882 and PHY/ASC93-18152 (arpa supplemented). J.M. also acknowledges a Fellowship (P.F.P.I.) from Ministerio de Educación y Ciencia of Spain.